\documentclass[aps,prd,preprintnumbers,showpacs,twocolumn,nofootinbib]{revtex4}
\usepackage{graphicx}

\usepackage{epsfig}
\usepackage{rotating}
\usepackage{dcolumn}
\usepackage{bm}

\newcommand{\PRE}[1]{}       

\def\gsim{\lower0.5ex\hbox{$\:\buildrel >\over\sim\:$}}
\def\lsim{\lower0.5ex\hbox{$\:\buildrel <\over\sim\:$}}

\newcommand{\beq}{\begin{equation}}
\newcommand{\eeq}{\end{equation}}
\newcommand{\barr}{\begin{array}}
\newcommand{\earr}{\end{array}}

\newcommand{\ev}{\; {\rm eV}}
\newcommand{\gev}{\; {\rm GeV}}

\begin{document}
\renewcommand{\thefootnote}{\fnsymbol{footnote}}

\title{
\PRE{\vspace*{2.0in}}
What does the $\nu_\mu$ oscillate into?
\PRE{\vspace*{.4in}}
}

\author{Debajyoti Choudhury$^{1,2}$}
\author{Anindya Datta$^2$\PRE{\vspace{.1in}}
}
\affiliation{{}$^1$Department of Physics and Astrophysics, University of Delhi,
  Delhi 110007, India\\
{}$^2$Harish-Chandra Research Institute, Chhatnag Road, Jhusi, 
Allahabad 211019, India.}
\begin{abstract}
The favoured resolution of the atmospheric neutrino anomaly involves 
an oscillation of the muon neutrino to a different state. 
Current experiments allow for the latter to contain a significantly 
large fraction of a non-standard flavour. We demonstrate how the next 
generation of experiments may take advantage of matter effects 
to resolve this issue. 
\end{abstract}

\maketitle 

A combination of several seminal experiments~\cite{superK, sno, chooz,
kamland, k2k} have left us with the inescapable conclusion that
neutrinos oscillate amongst themselves.  We do know that the
muon-neutrino ($\nu_\mu$) mixes almost maximally ($\sin^2
2\theta_{\mu3} > 0.92$ at 90\% C.L) with another species $\nu_3$
(distinct from the electron-neutrino $\nu_e$) and that the mass
splitting $|\delta m^2_{32}| \sim 2 \times 10^{-3}
\ev^2$. Furthermore, $\nu_e$ mixes with a combination of $\nu_\mu$ and
$\nu_\tau$ with a similarly large angle $\theta_{e2} \simeq 30^\circ$
but a far smaller mass splitting ($\delta m^2_{21} \sim 7 \times
10^{-5} \ev^2$). The remaining mixing angle $\theta_{13}$ is
constrained to be very small~\cite{chooz}. It is often touted that
neutrino oscillation physics is now poised to move into the domain of
precision studies, with the major remaining questions primarily
centering on unraveling the sign of $\delta m^2_{32}$, accurate
measurement of all the parameters and looking for possible $CP$
violation in this sector.

Several other questions remain though, not the least of which relates
to the identity of $\nu_3$. For one, this could be a linear
combination of $\nu_\tau$ and a sterile neutrino ($\nu_S$):
\beq
\nu_3 = c_\xi  \; \nu_\tau + s_\xi  \; \nu_S 
\eeq
where $c_\xi \equiv \cos \xi$ and $s_\xi \equiv \sin \xi$. As long as
$\theta_3 \equiv \theta_{\mu 3} \simeq \pi / 4$, the above choice
would reproduce the charged current (CC) data at both
Super-K~\cite{superK} and SNO~\cite{sno}.  While data already rules
out $s_\xi = 1$, note that $s^2 _\xi = 0.25 \,(0.35)$ is still allowed
at 90\%(99\%) C.L.~\cite{habig}. In fact,  $\nu_S$ could even be an
antineutrino (say, the $\bar \nu_\tau$)~\cite{lepton_violating} if
one would allow for a violation of the total lepton number. 
If, in addition, a helicity flip is enhanced (being proportional to 
$m_{\nu}$, this is normally suppressed) or a new interaction is relevant,
this could even produce an active $\bar\nu_\tau$, and 
thus induce wrong-sign CC events (analogous
oscillations of the form $\nu_e, \nu_\mu \to \bar \nu_e$ 
 have been looked for at the BEBC
experiment~\cite{Cooper-Sarkar:1981pb} with no evidence being found).
Unfortunately, none of the current or forthcoming experiments,
including {\sc opera}~\cite{opera} and {\sc icarus}~\cite{icarus},
are well-equipped for such measurements.

In this article, we seek to point out how passage through dense matter
may amplify the effects of a non-zero $s_\xi$ so that $CC$ events can
be used as a discriminator.  Since the refractive indices for
different neutrino beams may be altered to a significant degree
\cite{msw} by their interactions with matter, clearly the survival
probability $P_{\mu \mu}$ will display non-trivial matter effects for
$\nu_\mu \leftrightarrow \nu_S$ oscillations while it would not do so
for $\nu_\mu \leftrightarrow \nu_\tau$ oscillations. For an arbitrary
$s_\xi$, we will see that the resultant $P_{\mu \mu}$ would not
necessarily lie in between the two extreme cases. Rather, the latter has a
non-trivial dependence on $s_\xi$. To quantify this,
let us begin by restricting ourselves to the relevant three-neutrino
subspace namely $(\nu_\mu, \, \nu_\tau, \, \nu_S)$, characterized, in
general, by two mass (square)-differences $\Delta_{1,2}$.
The effective Hamiltonian can then be expressed as
\beq
{\cal H} = \left( 
           \barr{ccc}
           \delta_1 \, s_{3}^2 
            & - \delta_1 \, s_{3} \, c_{3} \, c_\xi
            & - \delta_1 \, s_{3} \, c_{3} \, s_\xi \\[1ex]
         {\cal H}_{12}
            & \delta_1 \, c_{3}^2 \, c_\xi^2 + \delta_2 \, s_\xi^2
            & c_\xi \, s_\xi \,  (c_{3}^2 \, \delta_1 - \delta_2)\\[1ex]
            {\cal H}_{13} 
            & {\cal H}_{23}
            & c_{3}^2 \, s_\xi^2 \, \delta_1 + c_\xi^2 \, \delta_2 
                  - V_{NC}
           \earr
           \right)
\eeq
where ${\cal H}$ is symmetric and $\delta_i \equiv \Delta_i / (2 \,
E_\nu)$.  $V_{NC} \simeq - G_F \, N_n /2 \sqrt{2}$, with $N_n$
denoting the neutron density in matter, denotes the relevant matter
effect.

Note that, for a non-zero $s_\xi$, matter effect arises even in the
absence of any mixing with the $\nu_e$. On the other hand, for $s_\xi
= 0$, all matter-induced effects (including
those in the $\nu_\mu$--$\nu_\tau$ sector) are proportional to the
size of the relevant mixing with $\nu_e$ and are understandably
small~\cite{barger_matter, maoki, winter, indu_murthy, raj,
choubey_roy, choudhury_datta}. The approximation of a three-generation
mixing scenario is, thus, an excellent one and has the virtue of
exhibiting the essence of the effect.

\begin{figure}[t]
\vspace*{2.5ex}
\centerline{\hspace*{-5em}
\epsfig{figure=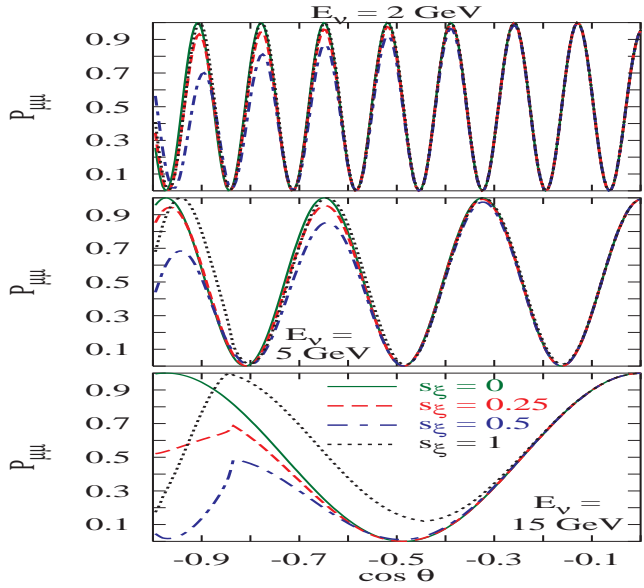,height=3.3in,width=4in,angle=0}}
\vspace*{-8ex}
\caption{\em The atmospheric $\nu_\mu$ survival probability as a
function of the incident angle.  Each panel refers to a different
$E_\nu$. The mixing angle $\theta_{\mu 3} = \pi / 4$
and $\Delta_1 = \Delta_2 = -0.003 \ev^2$.}
\label{figure:thetadep}
\end{figure}

Although the effective mixing angle now becomes a function of
instantaneous density, it should be realized that it is not
necessarily enhanced on account of the matter effect.  Rather, for
$s_\xi = 1$, when the $\nu_\tau$ decouples, the $\nu_\mu$--$\nu_S$
mixing, in vacuo, is close to maximal and the matter effect only
serves to decrease its size.  On the other hand, for small $s_\xi$, a
resonance enhancement of the $\nu_\mu$--$\nu_S$ mixing is indeed
possible.  In Fig.~\ref{figure:thetadep}, we exhibit the survival
probability for atmospheric $\nu_\mu$ of a fixed energy as a function
of its incident angle. For the numerical results, we have chosen
$\theta_{23} = \pi / 4$ and used the varying density profile of the
earth as given by the Preliminary Reference Earth Model (PREM)
\cite{prem}. The aforementioned suppression, for $s_\xi = 1$, is
clearly evinced by the corresponding $P_{\mu \mu}$ never vanishing
identically even at the minima (in stark contrast to $s_\xi = 0$). For
intermediate $s_\xi$, the two-generation simplification does not work,
and the change in $P_{\mu \mu}$ is more subtle. For small $E_\nu$,
where the diameter of the earth permits many oscillation periods, the
value of $P_{\mu \mu}$ at the maxima decrease monotonically. While
this effect scales with $s_\xi$ for small $\xi$, for large $s_\xi$,
the system quickly becomes rather nonlinear. For large $E_\nu$, the
matter effect is understandably much larger and non-linearity sets off
even for small $s_\xi$. 

\begin{figure}[!b]
\vspace*{0.ex}
\centerline{\hspace*{-5em}
\epsfig{figure=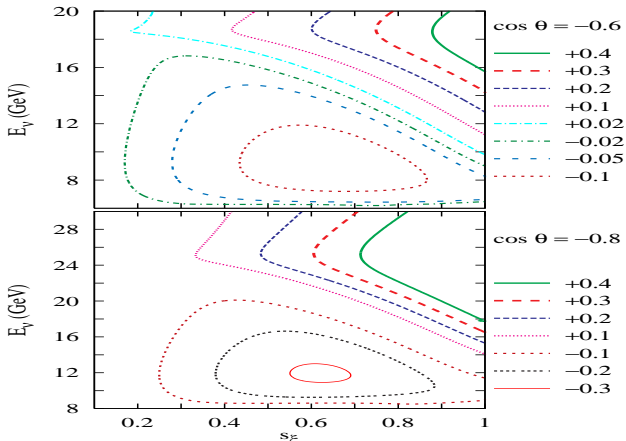,height=2.4in,width=3in,angle=0}}
\vspace*{-5ex}
\caption{\em Contours of constant $\cal R$ (defined in the text) in the $E_\nu
  - s_{\xi}$ plane. Parameters are  as in 
Fig.\protect\ref{figure:thetadep}.
}
\label{figure:contour}
\end{figure}

A detector-independent measure of the sensitivity is afforded 
by the quantity~\cite{choudhury_datta}
$
{\cal R}(E_\nu, \cos \theta,
s_{\xi}) \equiv \left[P_{\mu\mu} (s_{\xi}) - P_{\mu\mu} (0)\right]
/ \left[P_{\mu\mu} (s_{\xi}) + P_{\mu\mu} (0)\right]^{1/2}
$. 
The contours of constant ${\cal R}$ (Fig.\ref{figure:contour}), apart 
of illustrating the complicated dependence on $s_\xi$, also  
indicate the choices of $(E_\nu, \cos\theta)$  ideal to probe 
a given $s_\xi$. Note that high values of $|{\cal R}|$ are 
also obtainable at $E_\nu < 5 \gev$, but owing to the rapid oscillations
in $P_{\mu\mu}$, the exclusive use of such neutrinos would require 
very fine detector resolutions.

Such effects are characteristic of the scenario and can thus be used
even to measure $s_\xi$. Furthermore, it should be realised that with
$V_{NC}$ being of the opposite sign for antineutrinos, this deviation
from the simple $s_\xi = 0$ case would occur for only one of neutrinos
and antineutrinos.  Thus, if the detector were a magnetized one---such
as the proposed iron calorimeter (ICAL) at the India-based Neutrino
Observatory (INO)~\cite{ino}---and hence able to measure the
muon-charge, this could be used as a further check if any such
deviation were to be seen. Of course, whether $\nu_\mu$'s show this
effect or $\bar \nu_\mu$'s would depend on the sign of the mass
difference and thus would constitute a measure of the latter were
$s_\xi$ to be non-zero.  However, note that at large $E_\nu$, where the
difference is more pronounced, the atmospheric neutrino flux falls
rapidly. On the other hand, at lower energies, one would need both
very good energy resolution as well as angular resolution to draw any
positive inference.

\begin{figure}[!t]
\vspace*{-22ex}
\centerline{\hspace*{-5em}
\epsfig{figure=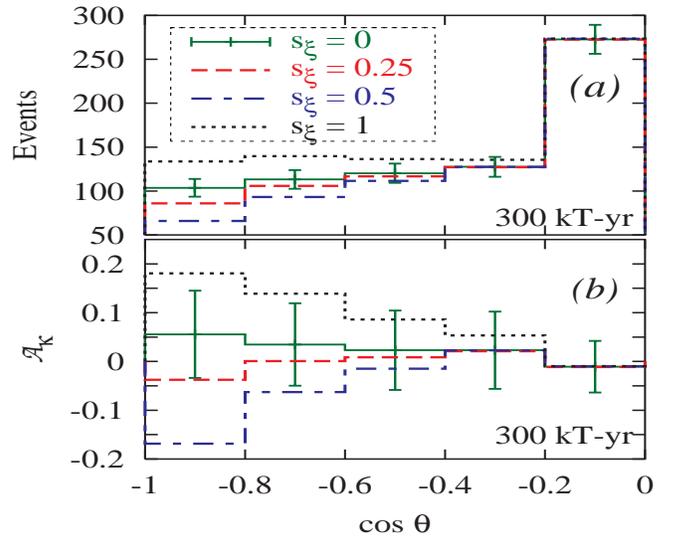,height=4in,width=4in,angle=0}}
\vspace*{-2ex}
\caption{\em {\em (a)} The number of $\nu_\mu$ CC events, 
as a function of incidence angle, 
from atmospheric neutrinos (with $5\gev < E_\nu < 40\gev$) 
in an ICAL \protect\cite{ino} like detector. 
The mixing angle $\theta_{\mu 3} = \pi / 4$
and $\Delta_1 = \Delta_2 = -0.003 \ev^2$.  {\em (b)} The 
weighted asymmetry between $\nu_\mu$ and $\bar \nu_\mu$ CC rates.
}
\label{figure:atmos}
\end{figure}

As an example, we consider the abovementioned ICAL, where the
projected energy threshold for muon detection is about 2~/gev and the
resolution is expected to be better than 0.5 /gev over the entire
range~\cite{ino,choudhury_datta}. In Fig.\ref{figure:atmos}$a$ we show
the number of $\mu^-$ events that such a detector would accrue from
atmospheric neutrinos~\cite{bartol} for a 300 kT-yr running period. 
To account for the threshold, as also the fact that
the difference is more apparent for larger energies, we integrate over
the range $E_\nu > 5 \gev$.  Clearly, the present limit of $s_\xi <
0.5$ would be easily distinguishable from the $s_\xi = 0$ case. In
fact, a sterile component one-fourth as small ($s_\xi = 0.25$) would
begin to be visible. Note that the $\bar\nu$ events are not expected
to show a similar effect for negative $\Delta_i$. It is thus
interesting to consider an asymmetry between the $\mu^\pm$ event rates
$N_\pm$
\[
{\cal A}_\kappa \equiv (N_- - \kappa \, N_+) 
                  / (N_- + \kappa \, N_+) \ .
\]
Using $\kappa \simeq 2.09$ to compensate for the difference in
$\nu_\mu$ and $\bar \nu_\mu$ CC cross-sections over the entire range,
the corresponding ${\cal A}_\kappa$ is displayed in
Fig.\ref{figure:atmos}$b$. Clearly, this too is a good discriminant,
though not as powerful as $N_-$ itself, owing to the smallness of
$N_+$.

Although the resolving power of an ICAL-like detector using
atmospheric neutrinos has been demonstrated above, it is still
contingent upon us to identify possible experiments with a greater
sensitivity, in particular to probe the case of positive $\Delta_i$.
Thus, it would be particularly useful if one could avail of a beam of
high energy neutrinos.  The latter is possible in the context of
accelerator experiments, especially a future facility such as a
super-beam or a neutrino factory wherein an intense beam of muons is
to be accelerated to a not too high energy and stored in a storage
ring with a straight section directed towards a neutrino
detector~\cite{nu_fac_tech}.  With muons decaying in this straight
section thereby producing high intensity neutrino beams (both of
electron-- and muon-types) that are highly collimated in the direction
of the decaying muons, such facilities have two additional advantages
over neutrino experiments with conventional neutrino beams (arising
from $\pi^\pm$ decay), namely ($i$) a precise knowledge of the
$\nu_\mu$ and $\bar \nu_e$ fluxes helps reduce the systematic errors;
($ii$) assuming a $\mu$-beam of, say, 20 /gev energy, such neutrinos
have, on the average, energies higher than those of conventional
neutrino beams thereby increasing the cross-sections at the
detector.

\begin{figure}[htb]
\vspace*{-0ex}
\centerline{\hspace*{-5em}
\epsfig{figure=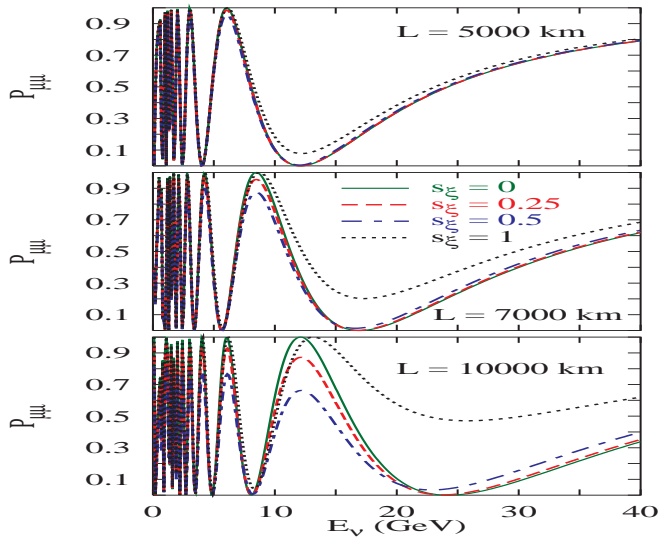,height=3.1in,width=4in,angle=0}}
\vspace*{-6ex}
\caption{\em The $\nu_\mu$ survival probability as a function of $E_\nu$
for different baselines.  Parameters are  as in 
Fig.\protect\ref{figure:thetadep}.
}
\label{figure:baseline}
\end{figure}

\begin{figure}[htb]
\vspace*{-0ex}
\centerline{\hspace*{-5em}
\epsfig{figure=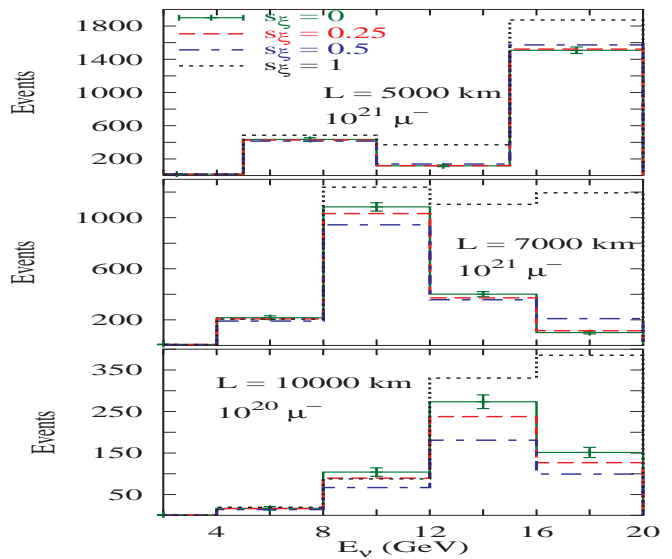,height=3.2in,width=4in,angle=0}}
\vspace*{-7ex}
\caption{\em The number of events expected 
for a 20 GeV muon storage ring and a 50 kT iron calorimeter with
a energy threshold of 2 GeV for the $\mu^\pm$~\protect\cite{ino}.
}
\label{figure:events_20gev}
\end{figure}

In Fig.~\ref{figure:baseline}, we display the survival probability for
such accelerator neutrinos as a function of their energy for a given
baseline. As expected, the difference between the two scenarios grows
with the baseline, for it is only with large baselines that the
neutrinos sample the larger densities in the earth's interior.  With
the differences starting to become apparent only for baselines larger
than 2000 km, a facility such as ICAL/INO~\cite{ino} would be ideally
placed to make such measurements particularly with neutrinos from JHF
(baseline $\sim$ 5000 Km), CERN ($\sim 7000$ km), or Fermilab ($\sim$
10000 Km).  For a given baseline, the difference between the scenarios
is expectedly more pronounced at larger neutrino
energies. Furthermore, larger $E_\nu$ translates to larger
cross-sections and, hence, larger event numbers. On the other hand,
the dependence of the neutrino flux on $E_\nu$ needs to be taken into
account. Convoluting all such effects, we show, in
Fig.\ref{figure:events_20gev}, the expected number of events for a
typical iron calorimeter detector~\cite{ino}. Although a larger
$E_\mu$ is preferable, both on account of larger cross-sections as
well as a tighter beam collimation, we choose to work with a likely
first-generation configuration, namely $E_\mu = 20 \gev$ and a
total of $10^{21}$ decaying muons.

%
Working with the assumption of $\theta_{e3} = 0$, we neglect, in
Fig.\ref{figure:events_20gev}, any {\em wrong sign} muon event (which,
anyway, can be distinguished by a magnetized detector) caused by the
$\bar \nu_e$ from $\mu^-$ decay oscillating into $\bar \nu_\mu$ while
traversing the earth.  For a baseline of 5000 km, the sensitivity is
marginal and $s_\xi = 0.5$ just about separable from $s_\xi = 0$. At
7000 km, on the other hand, one can easily explore down to $\sin \xi =
0.25$. For 10000 km, things improve dramatically and even $\sin \xi = 0.25$
is remarkably distinguishable with just one-tenth the number of 
muons (Fig.\ref{figure:events_20gev}).
A further countercheck is afforded by the fact that, for $\delta_i <
0$, one would expect a matter effect only for neutrinos (i.e., $\mu^-$
run) and not antineutrinos (a $\mu^+$ run). However, due to the lower
cross sections for $\bar \nu_\mu$, achieving similar significance in
the negative result would require twice the running time.

\begin{figure}[!t]
\vspace*{-2ex}
\centerline{\hspace*{-5em}
\epsfig{figure=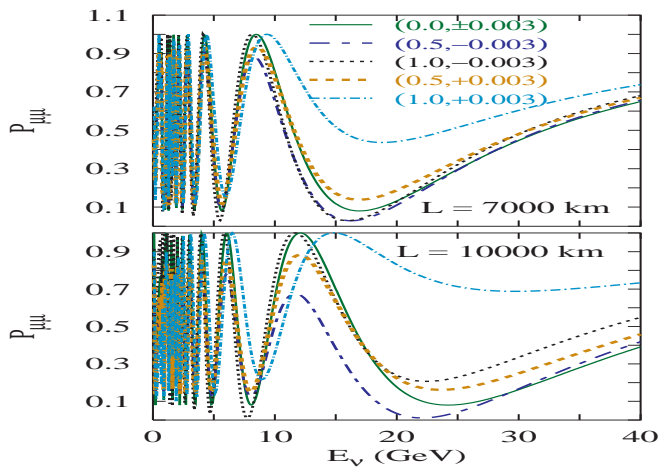,height=2.8in,width=4in,angle=0}}
\vspace*{-7ex}
\caption{\em The neutrino survival probability as a function of the
neutrino energy for different baselines and $\sin^2 2\theta_{\mu 3} =
0.92$.  The
parentheses in the legend refer to $(s_\xi, \Delta_i)$ combinations.
}
\label{figure:baseline_no_max}
\end{figure}

It should be realized that some of the details of
Figs.\ref{figure:thetadep}\&\ref{figure:baseline} are a consequence of
our choice of $\theta_{\mu 3} = \pi / 4$. As
Fig.\ref{figure:baseline_no_max} shows, were this mixing to be
non-maximal, $P_{\mu \mu}$ would be strictly positive for the
canonical case of $s_\xi = 0$.  On the other hand, for $s_\xi = 1$, at
resonance, the effective $\nu_\mu$--$\nu_S$ mixing would get enhanced,
thereby allowing $P_{\mu \mu}$ to reach zero at the minima. For
intermediate $s_\xi$, the two effects get intermixed leading to a
damping in the $P_{\mu \mu}$ oscillation at both ends (minima as well
as maxima).  Note that a non-maximal value of $\theta_{\mu3}$ renders
the sign of $\delta_i$ very important. Were $\delta_i$ to be positive
instead, clearly there would be no resonance enhancement of the
transition probability even for $s_\xi = 1$.

\begin{figure}[!t]
\vspace*{-1ex}
\centerline{\hspace*{-0em}
\epsfig{figure=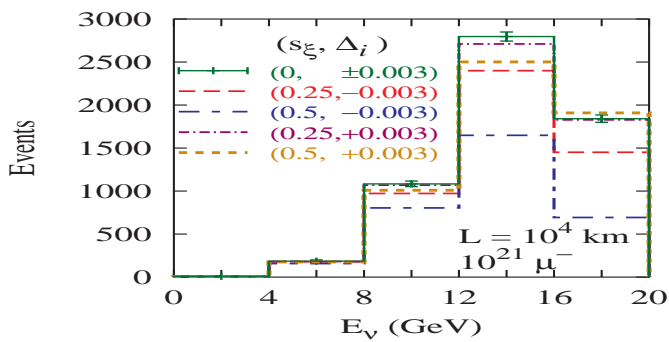,height=1.8in,width=3.8in,angle=0}}
\vspace*{-2ex}
\caption{\em As in Fig.\protect\ref{figure:events_20gev}, but for
$\sin^2 2\theta_{\mu 3} = 0.92$ instead.}
\label{figure:events_no_max}
\end{figure}

The corresponding event expectations are displayed in
Fig.\ref{figure:events_no_max}. Clearly, deviation of $\theta_{\mu 3}$
from maximality renders a non-zero value of $s_\xi$ even more visible,
at least for $\Delta_i < 0$. A positive value for the latter
understandably reduces the significance to the extent that,
even with $10^{21}$ muons, the case of $s_\xi = 0.25$ is just barely 
discernible. However, it should be kept in mind
that, in such a case, antineutrino oscillations have an enhanced
sensitivity and the effect would be clearly visible
 even accounting for a reduced cross section.

In summary, we have exhibited that matter effects may be used
profitably in resolving the vexing issue of the identity of the
neutrino state that $\nu_\mu$ oscillates into, especially since the
former is still allowed to contain a significantly large fraction of a
sterile state (or, equivalently, an antineutrino state).  Studies with
atmospheric neutrinos may themselves be used to obtain a reasonably
high degree of resolution. For example, with just 300 kT-yr exposure,
the planned ICAL/INO detector should be sensitive to a sterile
admixture one-fourth the size of current bounds. With the use of a
muon storage ring with a very long baseline, the resolving power
improves dramatically. Furthermore, such experiments are shown to be
quite sensitive to the deviation of the effective atmospheric neutrino
mixing angle from maximality thereby allowing for an accurate
measurement of the same. And while we have neglected the mixing with
$\nu_e$ in our analysis, such effects have been demonstrated to be
small and would not lead to any significant deterioration in the
analysing power.

\newcommand{\plb}[3]{{Phys. Lett.} {\bf B#1}, #2 (#3)}                  %
\newcommand{\prep}[3]{Phys. Rep. {\bf #1} #2 (#3)}                   %
\newcommand{\rpp}[3]{Rep. Prog. Phys. {\bf #1} #2 (#3)}             %
\newcommand{\np}[3]{Nucl. Phys. {\bf B#1} #2 (#3)}                     %
\newcommand{\npbps}[3]{Nucl. Phys. B (Proc. Suppl.)
           {\bf #1} #2 (#3)} %
\newcommand{\sci}[3]{Science {\bf #1} #2 (#3)}                 %
\newcommand{\zp}[3]{Z.~Phys. C{\bf#1} #2 (#3)}
\newcommand{\epj}[3]{Eur. Phys. J. {\bf C#1} #2 (#3)}
\newcommand{\mpla}[3]{Mod. Phys. Lett. {\bf A#1} #2 (#3)}             %
\newcommand{\jhep}[2]{{Jour. High Energy Phys.\/} {\bf #1} (#2) }%
\newcommand{\astropp}[3]{Astropart. Phys. {\bf #1}, #2 (#3)}            %
\newcommand{\ib}[3]{{ ibid.\/} {\bf #1} #2 (#3)}                    %
 \newcommand{\app}[3]{{ Acta Phys. Polon.   B\/}{\bf #1} #2 (#3)}%
\newcommand{\nuovocim}[3]{Nuovo Cim. {\bf C#1} #2 (#3)}         %
\newcommand{\yadfiz}[4]{Yad. Fiz. {\bf #1} #2 (#3);             %
Sov. J. Nucl.  Phys. {\bf #1} #3 (#4)]}               %
\newcommand{\jetp}[6]{{Zh. Eksp. Teor. Fiz.\/} {\bf #1} (#3) #2;
           {JETP } {\bf #4} (#6) #5}%
\newcommand{\philt}[3]{Phil. Trans. Roy. Soc. London A {\bf #1} #2
        (#3)}                                                          %
\newcommand{\hepph}[1]{hep--ph/#1}           %
\newcommand{\hepex}[1]{hep--ex/#1}           %
\newcommand{\astro}[1]{(astro--ph/#1)}         %

\end{document}